\documentclass[10pt]{article}
\usepackage{express}
\usepackage[detect-all]{siunitx}
\begin{document}

\title{Multipolar analysis of electric and magnetic modes excited by vector beams in core-satellite nano-structures}

\author{John Parker\authormark{1,2}, Stephen Gray\authormark{3}, Norbert Scherer\authormark{1,4,*}}
\address{\authormark{1}James Franck Institute,
\authormark{2}Department of Physics,
\authormark{4}Department of Chemistry, The University of Chicago, Chicago, IL 60637\\
\authormark{3}Center for Nanoscale Materials, Argonne National Laboratory, Chicago, IL 60439}

\email{\authormark{*}nfschere@uchicago.edu}

\begin{abstract}
    Core-satellite structures are known to exhibit magnetic modes at optical frequencies and their characterization is important for the development of metamaterials and metafluids. 
    We develop a finite-difference time-domain electrodynamics simulation and multipolar analysis approach and apply it to identify the electric and magnetic multipolar nature of modes excited in core-satellite structures composed of silver nanoparticles decorated on dielectric spheres.
    In addition to excitation with linearly polarized scalar beams, we investigate the scattering and multipolar properties induced by cylindrical vector beams.
    In contrast to linearly polarized beams, the nature of the polarization state in these beams (radial, azimuthal, or ``shear'') can selectively excite, enhance, and rotate a family of multipolar modes.
    Displacement currents induced in the nanoparticle gaps are investigated to better understand the nature of these excitations.
    We show that the efficiency of driving these modes depends on nanoparticle density and placement.
    We propose that selective magnetic and electric excitations can be codified as ``selection rules'' associated with the symmetries of the beams and particles.
\end{abstract}

\section{Introduction}


Creating negative-index metamaterials requires both a negative permittivity and permeability at the same excitation frequencies and this is tantamount to creating strong electric and magnetic resonances\cite{veselago1968electrodynamics,smith2004metamaterials,bourgeois2017self,fan2010self}.
Metamaterials and metafluids can potentially be built from suitable nanoscale structured ``meta-atom'' building blocks in a bottom-up fashion \cite{alu2006negative,sheikholeslami2013metafluid,fruhnert2014towards,qian2015raspberry,vallecchi2011collective,muhlig2011self,ponsinet2015resonant,gandra2012plasmonic}.
It is thus important to be able to measure and characterize the dipolar and higher-order multipolar resonances in nanoscale engineered materials by optical spectroscopic and numerical simulations for optimization. 
An exciting prospect is the selective excitation of electric and magnetic modes in such nanostructured materials by the use of vector beams\cite{uttam,das2015beam,wozniak2015selective}.
Quantitative comparison with simulation, which is critical for spectroscopic assignment and detailed understanding of the modes excited, requires consideration of the multipolar nature of the modes created with both scalar and vector beams of light, the angular scattering of radiation, and to be able to do so for arbitrary shaped nanostructures.
Moreover, controlling the selectivity and enhancement of multipolar modes with vector beams may aide in the design and application of nanoscale antennas used for remote sensing\cite{hancu2013multipolar,kujala2007multipole,oldenburg1999light}.

Theoretical and experimental studies of core-satellite nanostructures consisting of dielectric cores and metal nanoparticle shells have identified collective magnetic dipolar modes under linear plane wave excitation \cite{alu2006negative,sheikholeslami2013metafluid,fruhnert2014towards,qian2015raspberry,vallecchi2011collective,muhlig2011self,ponsinet2015resonant,gandra2012plasmonic}.
These studies used a method of multipolar analysis to identify the dipolar modes of the system.

Our new electrodynamics-multipolar analysis (ED-MA) method for simulating and analyzing multipolar mode excitation in core-satellite structures uses the finite-difference time-domain (FDTD) method \cite{taflove1995computational}, specifically a freely available software package \cite{OskooiRo10}, combined with a multipolar expansion of the scattered fields via a spherical monitor placed in the spatial grid of the simulation centered on the nanostructure to collect the scattered fields (see Fig. \ref{fig:setup}(a)).
This is used to compute multipoles up to 16 poles with high fidelity.
This method of multipolar analysis can be easily extended to any scattering structure, including arbitrary geometries and materials via standard FDTD software.

We demonstrate this rigorous ED-MA multipolar characterization method through application to dielectric core-Ag nanoparticle satellite structures under both scalar and vector beam illumination.
With the ED-MA method we can address in detail the types of magnetic modes that are created in core-satellite structures and ramifications of angular directional detection in experiments. 
Excitation with azimuthally polarized beams results in selective excitation of magnetic modes while radially polarized beams selectivity excite electric modes.
Specifically, we show that magnetic modes that arise from linearly polarized light rely on the finite size of the nanostructure and retardation whereas magnetic dipolar and quadrupolar modes excited with azimuthally polarized beams have no such dependence and are selectively excited, with high contrast vs. electric modes. 
The findings described in detail in this paper are in very good agreement with experiments \cite{uttam}.

Specifically, we study the effects of varying nanoparticle density and placement to understand and classify the optical modes of core-satellite structures.
The scattering spectra are found to change simply by rearranging the positions of the nanoparticles on the dielectric surface.
An ordered arrangement of nanoparticles can also spectrally shift the magnetic dipole mode of a disordered arrangement.
Displacement currents between silver nanoparticles are shown to be the driving factor of the collective magnetic modes.

Multipolar expansion analysis is also used to investigate the angular distribution of radiation from core-satellite structures illuminated by linearly polarized light and vector beams.
For a \SI{330}{nm} diameter SiO$_2$ core with a densely covered \SI{40}{nm} diameter Ag nanoparticle structure, the magnetic dipole and magnetic quadrupole modes are the dominant modes excited by linearly and azimuthally polarized light at $\SI{1.5}{eV}$, and are found to spatially interfere with one another.
By analyzing the angular distribution of these modes and interference patterns, we find a strong destructive interference in the backwards scattering direction.
These multipolar interference patterns are important to understand and consider when performing experiments that only collect scattered light over a finite range of angles.

The excitations created with other cylindrical vector beams (e.g. radially polarized and ``shear'' beams) are also presented. 
The variety of excitation fields and their selectivity allow us to identify a new class of ``selection rules'' for understanding the spectroscopy of nano-plasmonic, dielectric, and (nano-)metamaterials.
These selection rules can be utilized to target specific mode excitations and control the angular distribution of radiation, providing a new path to design and optimize the construction of new types of metamaterials.
Previous theoretical analysis has been reported that shows similar selective excitation properties via vector beams\cite{das2015beam}, but these methods are constrained to a single sphere as the sample.
By contrast, our approach is more general, capable of producing vector beams and performing a multipolar analysis of the scattered fields that works for a wider range of nanostructures and source conditions.




\section{Methods}

\subsection{Creating core-satellite structures}

Two types of core-satellite structures are considered: those with randomly positioned metal nanoparticles and those with an ordered arrangement.
To build a random arrangement, positions on the surface of the sphere (i.e. the SiO$_2$ dielectric core) are randomly and uniformly sampled by choosing two uniform random variables $\phi = U[0,2\pi]$ and $\tau = U[-1,1]$, and setting $\theta = \cos^{-1}(\tau)$.
If the radius of the dielectric core is $r_1$ and the radius of a nanoparticle is $r_2$, then a metal nanoparticle is placed at spherical coordinates $(r_1 + r_2, \theta, \phi)$.
The nanoparticle is placed only if it satisfies a minimum surface-to-surface separation distance of $d_{\text{sep}}$ between all other nanoparticles.
This process is repeated until $N_p$ nanoparticles have been placed.

We build ordered core-satellite structures via a minimum potential method, wherein each nanoparticle has a (fictitious) charge of $+1$. The potential energy of the system is
\begin{equation}
    U(\theta_i, \phi_i) = \sum_{i<j} \sum_{j=1}^{N_p}\: { \dfrac{1}{d_{ij}} }
    \label{eqn:ordered_core_satellite}
\end{equation}
where $d_{ij}$ is the distance between particle $i$ and particle $j$.
$U(\theta_i, \phi_i)$ is then minimized by gradient descent in $2N_p$--dimensional configuration space.
The computed spherical coordinates $(r_1 + r_2, \theta_i, \phi_i)$ are then used to create a core-satellite structure with an ordered nanoparticle arrangement.

These core-satellite structures are embedded in the FDTD grid using a sub-pixel smoothing algorithm provided by the MEEP software \cite{FarjadpourRo06,OskooiKo09}. 
The SiO$_2$ core is given a constant index of refraction $n_{\text{core}} = 1.46$, and the surrounding medium is oil of index $n_{\text{b}} = 1.51$.
The silver nanoparticles are modeled with a dispersive permittivity $\varepsilon_{\text{ag}}(\omega)$ and permeability $\mu_\text{ag} = 1$.
Experimental data of silver's dielectric function in the visible is used to fit $\varepsilon_{\text{ag}}(\omega)$ to a Drude-Lorentz model needed by the FDTD software\cite{johnson1972optical,rakic1998optical}.
All simulations were performed with a \SI{2}{nm} spatial resolution (i.e. cell voxels were \SI{2}{nm} on edge) and a Courant factor of $0.5$.
A $\SI{40}{nm}$ thick perfectly matched layer (PML) boundary condition was employed to model fields scattering to infinity.

Most of our modeling and analysis prsented below is for core-satellite structures where the dielectric core radius is $r_1 = \SI{165}{nm}$ and the Ag nanoparticle radius is $r_2 = \SI{20}{nm}$. 
The maximum number of nanoparticles used is $N_p = 162$.
For random clusters, we choose $d_\text{sep} = \SI{2}{nm}$ to prevent nanoparticles from touching.
We also consider the case where $d_\text{sep} = \SI{-2}{nm}$, allowing the particles to overlap by $\SI{2}{nm}$.
Similar conditions are considered for core radii of 82.5, 100, 125, 150, \SI{180}{nm} and $N_p = 40,59,92,132,190$, respectively.

\subsection{Producing vector beam sources in FDTD calculations}

\begin{figure}[ht!]
\centering\includegraphics{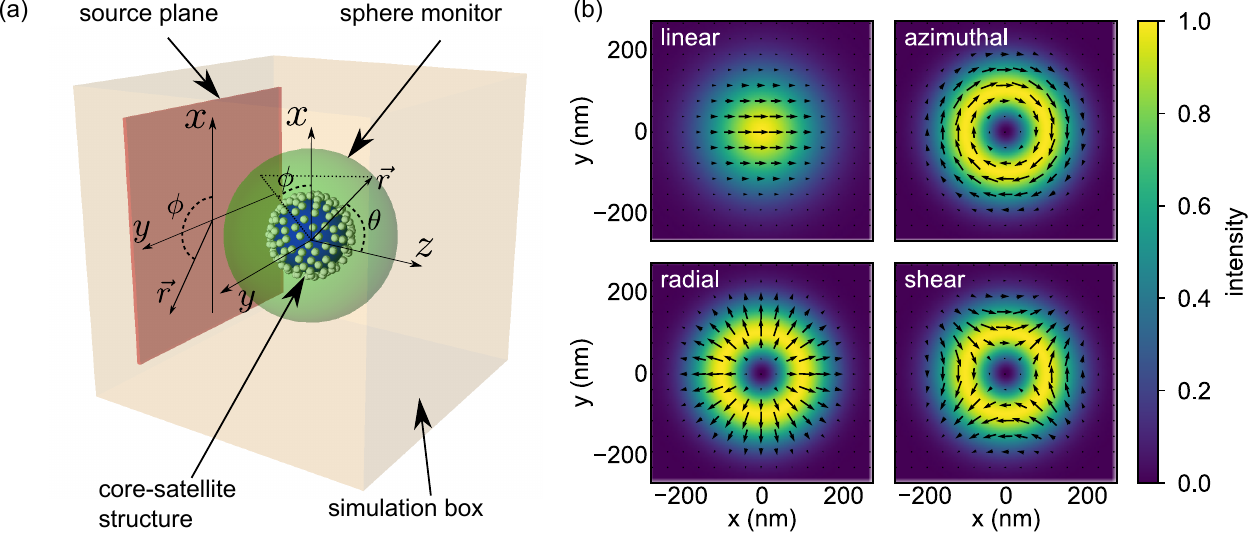}
\caption{Details of the FDTD simulations. (a) Simulation box, including a core-satellite scattering object, a spherical monitor to collect the Fourier transformed scattered fields, and a plane for the incident source.
    Perfectly matching absorbing layers (not explicitly shown) are placed in the outermost regions of all sides of the simulation box (b) The four different sources used: linear, radial, azimuthal, and shear beams. 
    The arrows indicate the instantaneous direction of electric field polarization in the source plane.}
    \label{fig:setup}
\end{figure}

The vector beams are created from linear combinations of the lowest-order Hermite-Gaussian modes \cite{zhan2009cylindrical,novotny2012principles}. Together with the linearly polarized beam, they are
\begin{align}
  \begin{split}
    \boldsymbol{E}_{\text{lin}} &= \text{HG}_{00} \boldsymbol{\hat x } \\
    \boldsymbol{E}_{\text{azi}} &= \text{HG}_{01} \boldsymbol{\hat x } - \text{HG}_{10} \boldsymbol{\hat y } \\
    \boldsymbol{E}_{\text{rad}} &= \text{HG}_{10} \boldsymbol{\hat x } + \text{HG}_{01} \boldsymbol{\hat y } \\
    \boldsymbol{E}_{\text{shear}} &= \text{HG}_{01} \boldsymbol{\hat x } + \text{HG}_{10} \boldsymbol{\hat y }
  \end{split}
  \label{vector_beam_fields}
\end{align}
These sources are generated in FDTD simulations using an electric surface current density $\boldsymbol{J}_E(x,y)$ in a source plane away from the scattering object (see Fig. \ref{fig:setup}).
The current sources for each beam are given the same spatial profile as the desired electric field based on the field equivalence principle \cite{rengarajan2000field},
\begin{align}
  \begin{split}
    \boldsymbol{J}_{\text{E,lin}} &= \text{HG}_{00} \boldsymbol{\hat x } \\
    \boldsymbol{J}_{\text{E,azi}} &= \text{HG}_{01} \boldsymbol{\hat x } - \text{HG}_{10} \boldsymbol{\hat y } \\
    \boldsymbol{J}_{\text{E,rad}} &= \text{HG}_{10} \boldsymbol{\hat x } + \text{HG}_{01} \boldsymbol{\hat y } \\
    \boldsymbol{J}_{\text{E,shear}} &= \text{HG}_{01} \boldsymbol{\hat x } + \text{HG}_{10} \boldsymbol{\hat y }
  \end{split}
  \label{vector_beam_currents}
\end{align}
These sources create vector beams focused at the source plane.
The beam propagates to the sample and diverges slightly at and across the sample.
For this reason, we choose the source plane to be as close as possible to the scattering object (about \SI{50}{\nano\meter} from the nanoparticle surface).

The coordinate system is chosen such that the $x$ and $y$-axes span the source plane and the $+z$-axis is along the direction of propagation.
In the source plane, $\phi$ denotes the angle subtended by a vector $\vec r$ and the $x$-axis (see Fig. \ref{fig:setup}(a)).

\subsection{Multipole analysis of scattered fields}

With the system and sources as described above, the FDTD method involves numerically propagating the electric and magnetic fields in time.
To obtain frequency-resolved information, the fields are Fourier transformed and various quantities such as the Poynting vector or flux can be computed from these transformed fields.
A traditional FDTD technique for measuring the scattering intensity is to use a box monitor centered on the scattering object that computes the scattered flux through each of the six faces, so that $F_\text{scat} = \sum_{i=1}^{6}{F_i}$.
A scattering intensity is obtained by normalizing the scattered flux by the incident flux of the light source
\begin{equation}
    I_{\text{scat}}(\omega) = \dfrac{F_\text{scat}(\omega)}{F_\text{inc}(\omega)}
    \label{eqn:box_flux}
\end{equation}
A scattering cross-section $\sigma_\text{scat}$ is obtained by multiplying the scattering intensity by the area of the incident flux monitor.
A scattering efficiency $Q_\text{scat}$ is then obtained by dividing $\sigma_\text{scat}$ by the physical cross-section of the scattering object, 
\begin{equation}
    Q_{\text{scat}}(\omega) = \dfrac{\sigma_\text{scat}(\omega)}{\pi r^2}
    \label{eqn:scat_efficiency}
\end{equation}
where $r$ is radius of the scattering object.
The box monitor is limited to this type of analysis.

As depicted in Fig. \ref{fig:setup}(a), we introduce a spherical monitor that can be used to perform the same technique as well as a multipolar analysis of the scattered modes and precise angular scattering calculations.
The scattered electric field $\boldsymbol{E}_\text{scat}(\omega, r, \theta, \phi)$ can be expanded into the vector spherical harmonics (VSH's)\cite{muhlig2011multipole,grahn2012electromagnetic}, and we summarize the important components here.
There are electric modes $\boldsymbol{N}_{l,m}(\omega, r, \theta, \phi)$ and magnetic modes $\boldsymbol{M}_{l,m}(\omega, r, \theta, \phi)$, where $l$ is the order of the mode ($l = 1,2,3,\ldots)$ and $m$ is the orientation of the mode ($m = -l,-l+1, \ldots, 0, \ldots, l-1, l$). 
The scattered field is
\begin{equation}
    \boldsymbol{E}_{\text{scat}} (\omega, r, \theta, \phi) = \sum_{l=1}^{\infty}
    \sum_{m=-l}^{m=l}{a_{lm}(\omega) \boldsymbol{N}_{lm} + b_{lm}(\omega) \boldsymbol{M}_{lm}}
    \label{eqn:vcs_expansion}
\end{equation}
where $a_{lm}(\omega)$ and $b_{lm}(\omega)$ are the complex multipolar coefficients corresponding to the electric and magnetic modes, respectively, of order $l$ and orientation $m$.

The VSH's form an orthogonal inner product space, with orthogonality condition
\begin{align}
    \begin{split}
        \oint_\Omega \boldsymbol{N}_{lm} \cdot \boldsymbol{N}_{l^\prime m^\prime} \:d \Omega &= \delta_{ll^\prime}\delta_{m m^\prime}  \\
        \oint_\Omega \boldsymbol{M}_{lm} \cdot \boldsymbol{M}_{l^\prime m^\prime} \:d \Omega &= \delta_{ll^\prime}\delta_{m m^\prime}  \\
    \oint_\Omega \boldsymbol{N}_{lm} \cdot \boldsymbol{M}_{lm} \:d \Omega &= 0 
    \end{split}
    \label{vcs_orthogonallity}
\end{align}
where $\Omega$ is the surface of a sphere of radius $r_0$ and $\delta_{l l^\prime}$ is the Kronecker delta.
The multipolar coefficients $a_{lm}(\omega)$ and $b_{lm}(\omega)$ can be determined by combining  Eq.(\ref{eqn:vcs_expansion}) and Eq.(\ref{vcs_orthogonallity}):
\begin{align}
    \begin{split}
        a_{lm}(\omega) &= \oint_{\Omega} \boldsymbol{E}_{\text{scat}} (\omega, r = r_0, \theta, \phi) \cdot \boldsymbol{N}_{lm} (\omega, r = r_0, \theta, \phi) \:d \Omega \\
        b_{lm}(\omega) &= \oint_{\Omega} \boldsymbol{E}_{\text{scat}} (\omega, r = r_0, \theta, \phi) \cdot \boldsymbol{M}_{lm} (\omega, r = r_0, \theta, \phi) \:d \Omega
    \end{split}
    \label{eqn:vcs_decomposition}
\end{align}
Eq.(\ref{eqn:vcs_decomposition}) is the multipole decomposition. 
We choose a spherical coordinate system centered on the core-satellite structure and a spherical monitor of radius $R_\text{monitor}$.
The angles $\theta$ and $\phi$ (see Fig. \ref{fig:setup}(a)) are used to span the full $4\pi$ steradians ($\mathrm{sr}$) spherical monitor surface.
Eq.(\ref{eqn:vcs_decomposition}) can be used to numerically compute the multipolar scattering coefficients by collecting the complex-valued Fourier transformed fields $\boldsymbol{E}_{\text{scat}}$ on the surface of the spherical monitor.

To obtain the scattered flux per multipole, we make use of the following relation \cite{muhlig2011multipole}
\begin{equation}
    F_\text{scat}(\omega) = k^2 \sum_{l=1}^{\infty} \sum_{m=-n}^{n} {l(l+1)\left( |a_{lm}(\omega)|^2 + |b_{lm}(\omega)|^2 \right)}
    \label{eqn:multipole_flux}
\end{equation}
where $k=2\pi n_\text{b}/\lambda$.
Each $(n,m)$ term in the sum provides the scattered flux of the corresponding multipolar mode.
The sum of all such scattering fluxes gives the total scattered flux, $F_\text{scat}$, which are then normalized by the incident flux (see Eq.(\ref{eqn:box_flux})).

This method may be used to investigate angular distributions of radiation and the interference of different multipolar modes.
For instance, an electric mode of order $(l,m)$ and a magnetic mode of order $(l^\prime, m^\prime)$ will spatially interfere with one another.
Therefore, the angular scattering intensity is proportional to 
\begin{align}
    \begin{split}
        |a_{lm}\boldsymbol{N}_{lm} + b_{l^\prime m^\prime}\boldsymbol{M}_{l^\prime m^\prime}|^2 =& 
        \overbrace{|a_{lm}|^2|\boldsymbol{N}_{lm}|^2}^\text{electric mode} 
        + \overbrace{|b_{l^\prime m^\prime}|^2|\boldsymbol{M}_{l^\prime m^\prime}|^2}^\text{magnetic mode} \\
        &+ \underbrace{a_{lm}{b^*}_{l^\prime m^\prime}\boldsymbol{N}_{lm} \cdot {\boldsymbol{M}^*}_{l^\prime m^\prime}
        + {a^*}_{lm}{b}_{l^\prime m^\prime}{\boldsymbol{N}^*}_{lm} \cdot {\boldsymbol{M}}_{l^\prime m^\prime}}_\text{electric -- magnetic interference terms}
    \end{split}
    \label{eqn:interference}
\end{align}
These interference terms are calculated for any two modes of interest.
In the case of more than two modes scattering simultaneously, Eq.(\ref{eqn:interference}) can be generalized to multiple interference terms.

In addition to providing the multipolar analysis, the spherical monitor can also be used to determine the angular scattering distribution of radiation.
The Poynting vector of the scattered fields is $ \boldsymbol{S} = \boldsymbol{E}_\text{scat} \times \boldsymbol{H}_\text{scat}$.
The scattered flux can then be determined over any angular region of the sphere, $\Sigma$, by integrating $\boldsymbol{S}$ over that part of the sphere
\begin{equation}
    F_{\text{scat}}(\omega) = \oint_{\Sigma}{\boldsymbol{S}(\omega) \cdot \boldsymbol{\hat r} \:d \Sigma}
    \label{eqn:anguler_scattered_flux}
\end{equation}
We are also interested in examining the scattering in the forward or backward direction over some range of collection angles for comparison with experimental studies that may be limited in such a manner.
Choosing $\Sigma$ to be this surface, Eq.(\ref{eqn:anguler_scattered_flux}) can be expressed as
\begin{equation}
    F_{\text{scat}}(\omega) = \int_{\theta_{\text{min}}}^{\theta_{\text{max}}} 
                      \int_{0}^{2\pi} {
                      {\boldsymbol{S}(\omega) \cdot \boldsymbol{\hat r}\: r^2 \sin{\theta} \: d\phi d\theta} }
    \label{eqn:anguler_scattered_flux_specific}
\end{equation}
Eq.(\ref{eqn:anguler_scattered_flux_specific}) is computed numerically using the scattered fields from the sphere monitor at each frequency.

\section{Results}

\subsection{Nanoparticle density affects the scattering spectra}

\begin{figure}[ht!]
    \centering\includegraphics{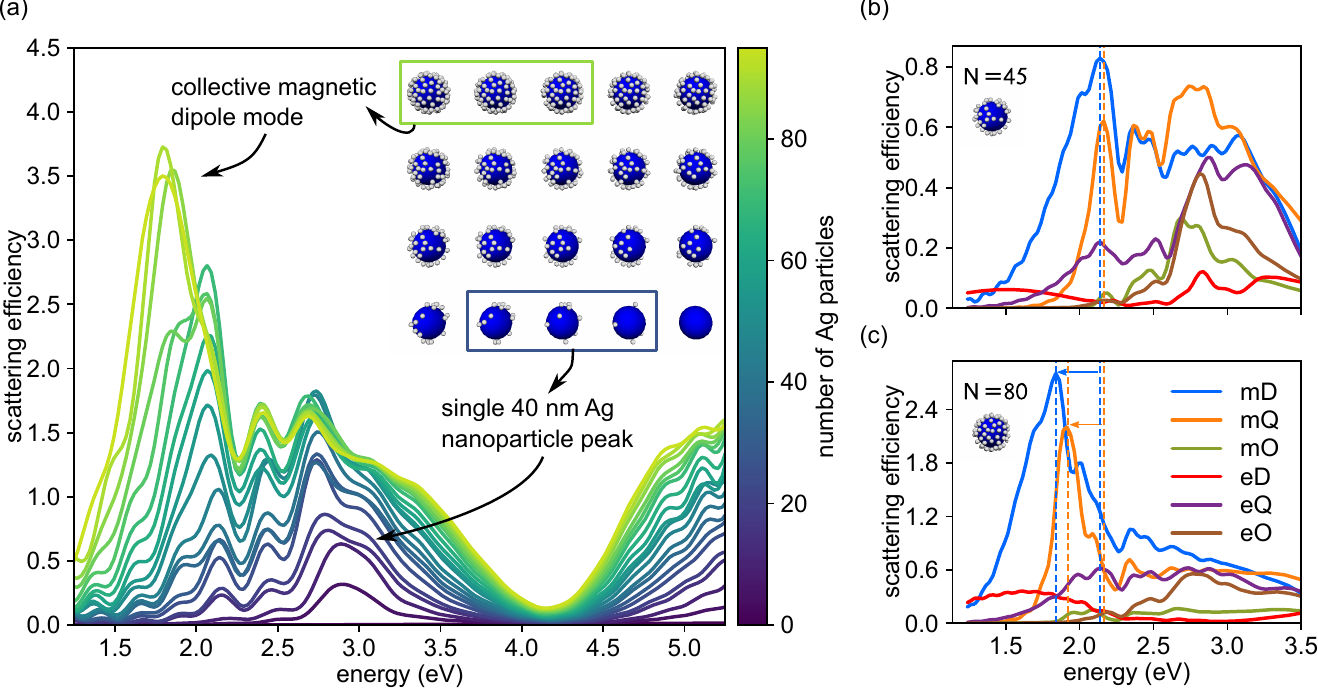}
    \caption{(a) Scattering intensity by linearly polarized beam excitation of core-satellite structures with increasing nanoparticle coverage.
             The \SI{2.9}{eV} peak corresponds to the single dipole resonance of a \SI{40}{nm} Ag nanoparticle.
             The \SI{1.8}{eV} magnetic dipole peak emerges at high nanoparticle coverage.
         (b) Multipole scattering for $N=45$ nanoparticles.
         (c) Multipole scattering for $N=80$ nanoparticles.
         This identifies the \SI{1.8}{eV} peak in (a) as a magnetic mode.
     The magnetic dipole and magnetic quadrupole modes in (c) red-shift $\SI{0.3}{eV}$ relative to (b).}
    \label{fig:varying_density}
\end{figure}

We simulated the scattering from core-satellite structures of increasing nanoparticle density to understand the spectroscopy of the core-satellite system and assign identities to spectroscopic features.
The progression of some features also give insight into the collective magnetic modes being excited in the core-satellite structure.
We begin with 5 silver nanoparticles (\SI{40}{nm} diameter) placed randomly on the surface of a dielectric core of radius \SI{120}{nm} and repeatedly add 5 additional nanoparticles until 95 nanoparticles were placed, which corresponds to roughly maximum random packing density.
A scattering spectrum for each of these structures is shown in Fig. \ref{fig:varying_density}(a).

The major scattering feature for low particle densities (5 to 15 nanoparticles) occurs at \SI{2.9}{eV}. 
This mode corresponds to the plasmon resonance of individual \SI{40}{nm} Ag spheres since the nanoparticles are spatially separated enough to minimize near-field interaction.
As the nanoparticle density increases (from 20 to 80 nanoparticles), features (modes) emerge at \SI{2.2}{eV} and \SI{2.5}{eV} due to near-field interactions.
A new mode at \SI{1.8}{eV} emerges at the highest densities (85-95 nanoparticles).
This is the magnetic mode of the core-satellite that only occurs when the density is large enough that near-field interactions can generate a collective excitation around the core perimeter.

These assignments stem from a multipolar analysis we did for two of the various clusters, one with $N_p = 45$ (Fig. \ref{fig:varying_density}(b)) and one with $N_p = 80$ (Fig. \ref{fig:varying_density}(c)).
At the lower nanoparticle density, magnetic dipole and magnetic quadrupole modes are present and have a peak around \SI{2.2}{eV} and additional features at higher energy.
The intensity of these modes is slightly larger than electric quadrupole and electric octupole modes present at higher energies.
At the higher nanoparticle density, the two magnetic modes red-shift by about \SI{0.3}{eV} relative to what was observed in the lower density structure.
The intensity of these magnetic modes are enhanced by about a factor of 4 relative to the lower density structure, while the electric modes remain of similar magnitude.
Thus, the emergence of the magnetic mode at \SI{1.8}{eV} can be understood as a magnetic dipole--magnetic quadrupole pair that red-shifts and intensifies rapidly with increasing nanoparticle density.

\subsection{Nanoparticle order and placement affects the scattering spectra}

\begin{figure}[ht!]
    \centering\includegraphics{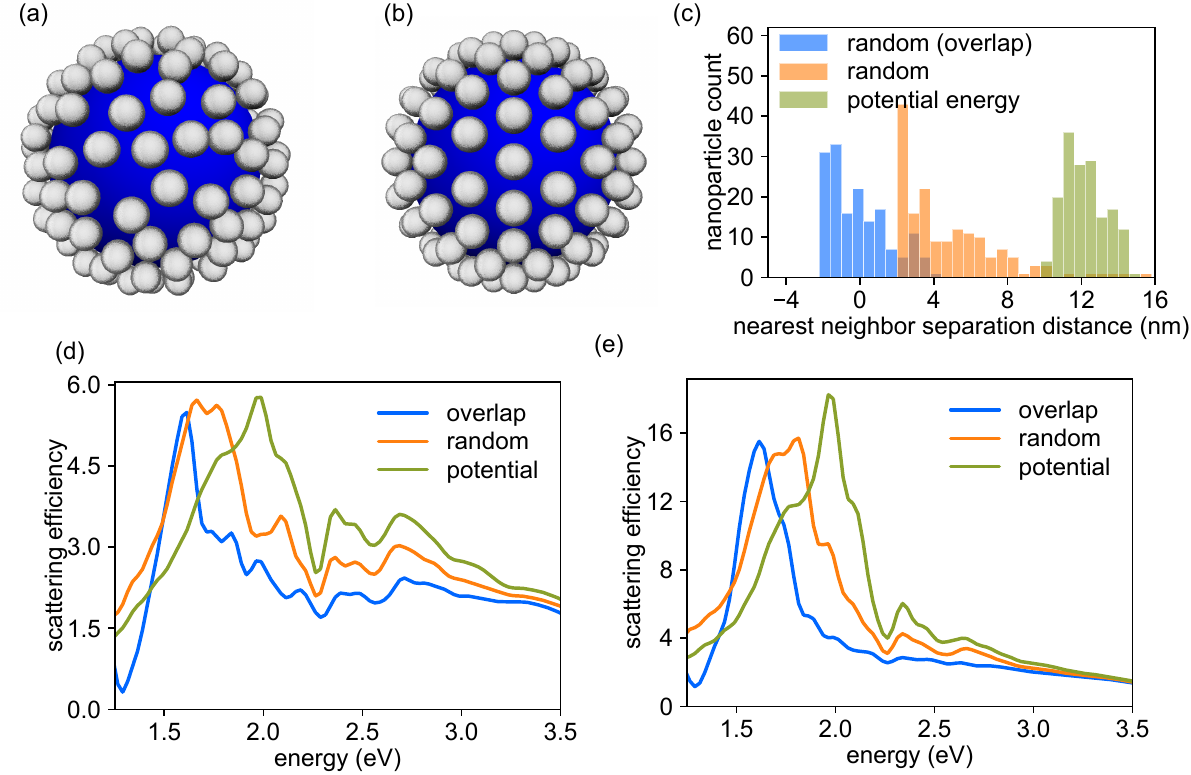}
    \caption{
        Spectroscopic effects of Ag nanoparticle packing.
        A core-satellite with $N_p = 162$ particles and core radius $r_1 = \SI{165}{nm}$ are created with (a) a random packing and (b) an ordered packing.
        (c) Distributions of nearest neighbor separations (surface to surface) for each method of creating core-satellites.
        (d,e) Scattering intensity of each type of core-satellite structure by a linearly polarized beam (d), and azimuthally polarized beam (e).}
    \label{fig:ordered}
\end{figure}

The magnetic mode can be understood as a net circulation of displacement current around the core-satellite that only occurs at high nanoparticle density. 
Therefore, the precise locations and ordering of the nanoparticles should affect the scattering properties.
Fig. \ref{fig:ordered}(a) shows a typical core-satellite structure created via random placement of the nanoparticles and
Fig. \ref{fig:ordered}(b) shows a core-satellite structure of the same size and nanoparticle density, but with an ordered arrangement of particles created via the minimum potential method (see Eq.(\ref{eqn:ordered_core_satellite})).

Three different nanoparticle placement algorithms were considered.
In the random overlap algorithm, nanoparticles are randomly placed on the surface of the sphere, and a small overlap of $d_\text{sep} = \SI{-2}{nm}$ is allowed between nanoparticles.
The random (non-overlap) algorithm uses the same total scattering analysis, but forces a minimum separation of $d_\text{sep} = \SI{2}{nm}$ between the nanoparticles.
The potential energy algorithm (see Eq.(\ref{eqn:ordered_core_satellite})) is used to create a maximally separated and ordered distribution of nanoparticles.

A useful statistic to distinguish different nanoparticle arrangements is the distribution of nearest neighbor separations of the nanoparticles.
The distribution of nearest neighbor distances for the different core-satellite creation methods is shown in Fig. \ref{fig:ordered}(c).
Random arrangements have an average separation about \SI{10}{nm} less than that of an ordered arrangement.

These differences in nanoparticle arrangement are manifest in the scattering spectra for linearly and azimuthally polarized beam excitation (Fig. \ref{fig:ordered}(d) and (e), respectively).
Ordered arrangements with larger separations have their magnetic modes blue-shifted by about \SI{0.3}{eV} relative to the magnetic mode in a random cluster.
A random arrangement of nanoparticles that allows \SI{2}{nm} overlap is red-shifted by about \SI{0.1}{eV} relative to an arrangement that prevents particle overlap.

\begin{figure}[ht!]
    \centering\includegraphics{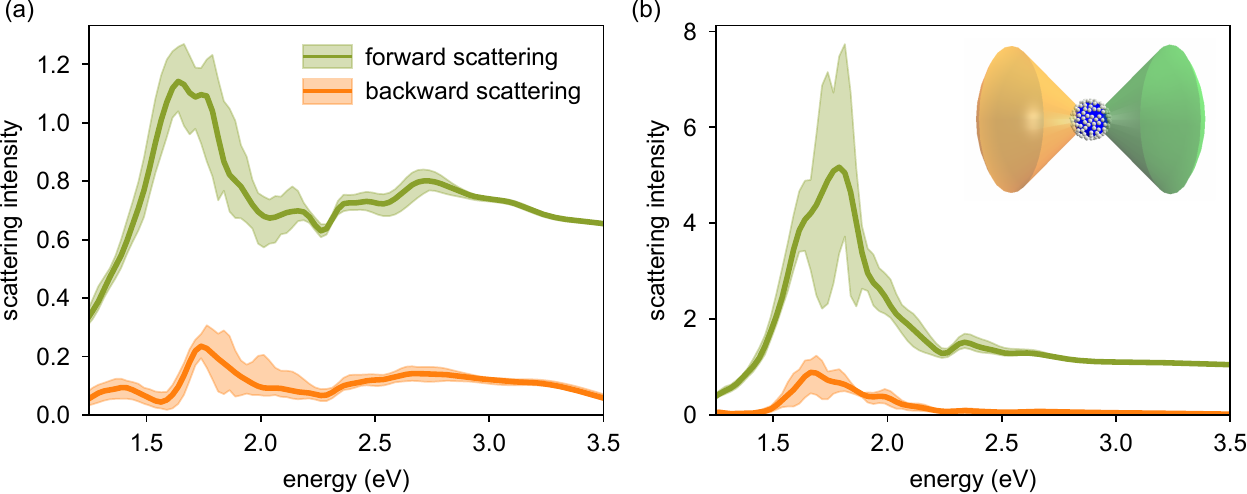}
    \caption{Forward and backward scattering by ten different permutations of the SiO$_2$-Ag NP core-satellite for (a) linearly polarized beam and (b) azimuthally polarized beam.
        The solid line is the average of the ten permutations, while the shaded region encompasses the region between the lower and upper bounds across all ten permutations.
             In both cases, back scattering is weaker than forward scattering by about a factor of $6$.}
    \label{fig:permutation}
\end{figure}

We also consider different random permutations of the nanoparticles on the surface for a fixed core size and nanoparticle density. 
Here, a permutation means a different arrangement of the same number of nanoparticles on the core and can be achieved by repeating the random placement algorithm.
Forward and backward scattering spectra for ten different random permutations are shown in Fig. \ref{fig:permutation} by linearly polarized beam excitation (a) and azimuthally polarized beam excitation (b).
The solid lines show the average of the spectra for all ten permutations, while the shaded regions encompass all ten spectra.
The angular scattering spectra are obtaining by using Eq.(\ref{eqn:anguler_scattered_flux_specific}) with $\theta_\text{min} = 0$ and $\theta_\text{max} = \pi/4$ for forward scattering and $\theta_\text{min} = 3\pi/4$ and $\theta_\text{max} = \pi$ for backward scattering.

The ten permutations result in slightly different scattering spectra for forward and backward scattering.
In addition, there are also features in the back scattering that are not apparent in the forward scattering.
For instance, a small peak at \SI{1.4}{eV} is present in the back scattering spectrum but not in the forward scattering spectrum for linearly polarized beam excitation.
This is an important consideration when comparing a simulation result with scattering spectra measured in experiments that can only collect scattered light over a finite range of angles.
The appearance of this peak can be explained by the angular distribution in Fig. \ref{fig:selective}(d); the magnetic dipole mode excited with linearly polarized light has maximum intensity in the $-z$ direction whereas the magnetic quadrupole mode has maximum intensity in the $\pm x$ direction.
Thus, if the scattering intensity is collected in a way that favors the $-z$ axis, the magnetic dipole mode will appear enhanced in the back scattering spectra as compared to the total scattering spectra.

\subsection{Understanding the magnetic dipole mode as a collective excitation}

As shown in Fig. \ref{fig:selective}(c), the linearly and azimuthally polarized beams are the only sources that are able to excite the magnetic dipole mode of the core-satellite structure.
In Fig. \ref{fig:selective}(d), these two modes are shown to have different orientations: the linearly polarized beam excitation has a $y$-oriented magnetic dipole and the azimuthally polarized beam excitation has a $z$-oriented magnetic dipole.
Although the multipolar analysis is able to provide this information, it does not provide an underlying mechanism for why these two beams excite the magnetic dipole mode nor explain the difference in orientation.

\begin{figure}[ht!]
    \centering\includegraphics{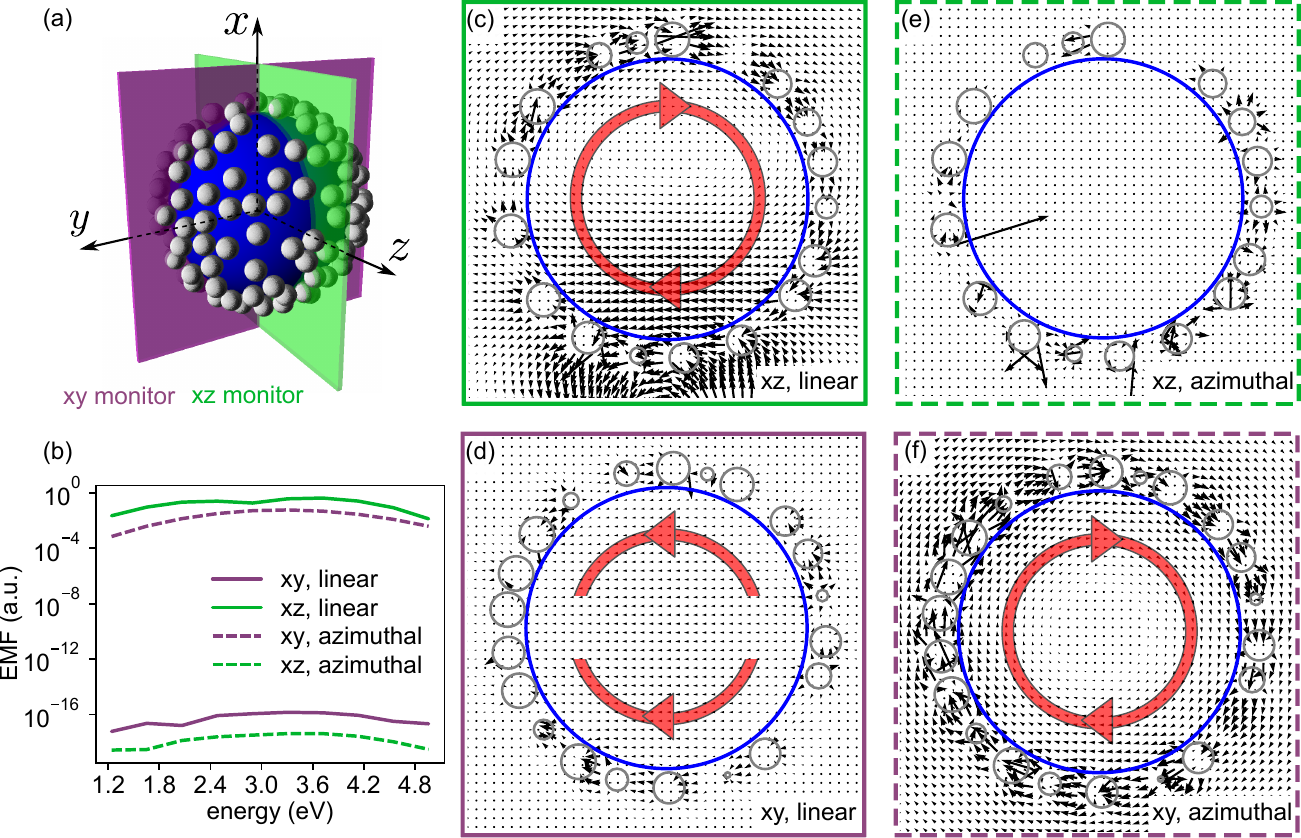}
    \caption{Visualizing the magnetic dipole modes excited by linearly and azimuthally polarized beams.
            (a) Two monitors are used to collect the Fourier transformed scattered near-fields inside cross-sections of the core-satellite.
            (b) Driving EMF from $\boldsymbol{E}_\text{inc}$ in each plane for both beams.
            (c-f) Total electric field inside the core-satellite structure in the $xz$ and $xy$ planes at \SI{1.5}{eV} for linearly and azimuthally polarized beams.
            (c) and (f) correspond to a circulation of displacement current around the circumference of the core-satellite.
            (d) and (e) do not display a net circulation of displacement current.}
    \label{fig:magnetic_mode_2}
\end{figure}

A more complete understanding is provided by the near-field electric fields in two cross-sections of the core-satellite structure: one in the $xy$ plane and one in the $xz$ plane (Fig. \ref{fig:magnetic_mode_2}(a)). 
The linearly or azimuthally polarized beams are incident along the $z$-axis.
A driving ``EMF'' is defined for each of the two planes as $\text{EMF} = \oint_\Omega{\boldsymbol{E}_\text{inc}} \cdot d \boldsymbol{l}$, 
where $\Omega$ is a closed loop through the nanoparticle surface, $\boldsymbol{E}_\text{inc}$ is the incident electric field, and $ d \boldsymbol{l}$ is a vector tangent to the loop.
The driving EMF in the cross-sections for excitation with linearly and azimuthally polarized beams is shown in Fig. \ref{fig:magnetic_mode_2}(b).
These driving EMFs suggest that linearly polarized light can generate a circulation of displacement current in the $xz$ plane and azimuthally polarized light can generate a circulation in the $xy$ plane.

This is further demonstrated by looking at the total field $\boldsymbol{E}_\text{tot} = \boldsymbol{E}_\text{inc} + \boldsymbol{E}_\text{scat}$ in each cross-section for the two beams at energy $\SI{1.5}{eV}$.
Fig. \ref{fig:magnetic_mode_2}(c) shows a circulating displacement current in the $xz$ plane due to the linearly polarized beam.
Fig. \ref{fig:magnetic_mode_2}(f) shows a similar circulating displacement current in the $xy$ plane due to the azimuthally polarized beam.
The arrow plots in Fig. \ref{fig:magnetic_mode_2}(c-d) and \ref{fig:magnetic_mode_2}(e-f) are each on the same scale.

When looking at the fields in the $xy$ plane for the linearly polarized beam (Fig. \ref{fig:magnetic_mode_2}(d)), there are two weaker excitations at the top and bottom halves of the core-satellite. 
However, these two excitations oppose each other, and the net result is zero circulation in this plane.
These displacement currents are likely responsible for the electric dipolar or quadrupolar modes presented and discussed in section \ref{sec:selective} below.
Similarly, the excitation created by the azimuthally polarized beam in the $xz$ plane is much weaker and doesn't appear to have any net circulation.

These results demonstrate an important difference between the magnetic dipolar modes excited by linearly and azimuthally polarized beams.
Although they have very similar spectral features, the underlying mechanisms of excitation are very different.
The azimuthally polarized beam is able to excite a circulation of displacement current due to its unique polarization in the beam's cross-section.
The linearly polarized beam does not have this capability, so the same magnetic mode cannot be excited by it.
However, the linearly polarized beam is able to excite a very similar mode in the $xz$ plane due to retardation effects along the $z$-axis. 
In particular, since the light is polarized in the $x$ direction, there is a phase lag along the $z$-axis due to retardation and a circulation of displacement current arises in the $xz$ plane (Fig. \ref{fig:magnetic_mode_2}(c)).

\begin{figure}[ht!]
    \centering\includegraphics{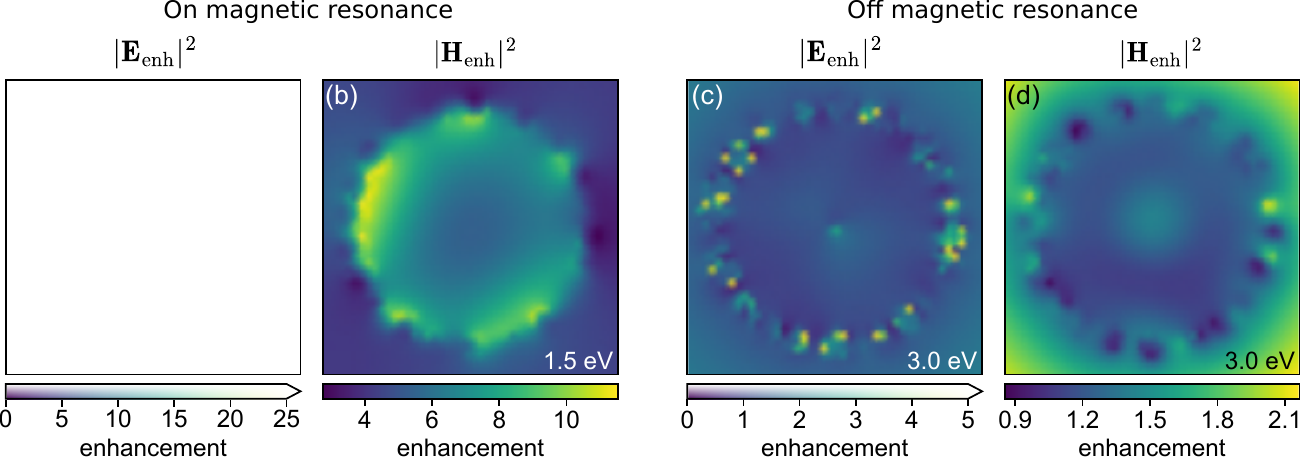}
    \caption{The magnetic dipole resonance is visualized on and off resonance of the magnetic mode for a core-satellite structure with an azimuthally polarized beam excitation. 
    (a) Electric field enhancement in an xy slice of the core-satellite on resonance ($\SI{1.5}{eV}$) of the magnetic mode. Strong enhancements are found in the gaps between nanoparticles.
    (b) Magnetic field enhancement on resonance of the magnetic mode. Strong enhancements are found inside the core region.
    (c) Electric field enhancement in an xy slice of the core-satellite off resonance ($\SI{3.0}{eV}$) of the magnetic mode. Enhancement is localized at the nanoparticle surfaces.
    (d) Magnetic field enhancement off resonance of the magnetic mode. Enhancement occurs outside of the core.}
    \label{fig:field_enhancements}
\end{figure}

To further demonstrate the presence of a magnetic mode with an azimuthally polarized beam, we look at the electric and magnetic field enhancements on resonance (\SI{1.5}{eV}) and off resonance (\SI{3.0}{eV}) of the magnetic mode.
Fig. \ref{fig:field_enhancements}(a,b) shows the electric and magnetic field enhancement on resonance in the $xy$ plane passing through the center of the core-satellite structure.
The electric field is strongly enhanced in the region of nanoparticle gaps and in the center of the core.
The electric field enhancement at the very center results because the cylindrical vector beams have vanishing intensity (of the incident beam) along the propagation axis which distorts the normalization.
The magnetic field enhancement is strong inside the core region and weak outside of it.
These field patterns are characteristic of a magnetic mode generated by a circulation of current around the core.

Away from the resonance of the magnetic mode, the electric field enhancement weakens in the nanoparticle gap regions and is instead localized to the surfaces of the nanoparticles (Fig. \ref{fig:field_enhancements}(c)).
The magnetic field enhancement is now expelled from the core region and stronger outside the cluster (Fig. \ref{fig:field_enhancements}(d)).
From the scattering spectrum shown in section \ref{sec:selective} for the azimuthally polarized beam, we see that at \SI{3.0}{eV} the magnetic dipole is much weaker compared to the higher order multipolar terms, including a magnetic hexadecapole term.
These higher order terms cause the electric field enhancement to be localized to the nanoparticle surfaces and expel the magnetic field from the core.

\subsection{Multipolar interference and the affects of size variation of the core diameter}

\begin{figure}[ht!]
    \centering\includegraphics{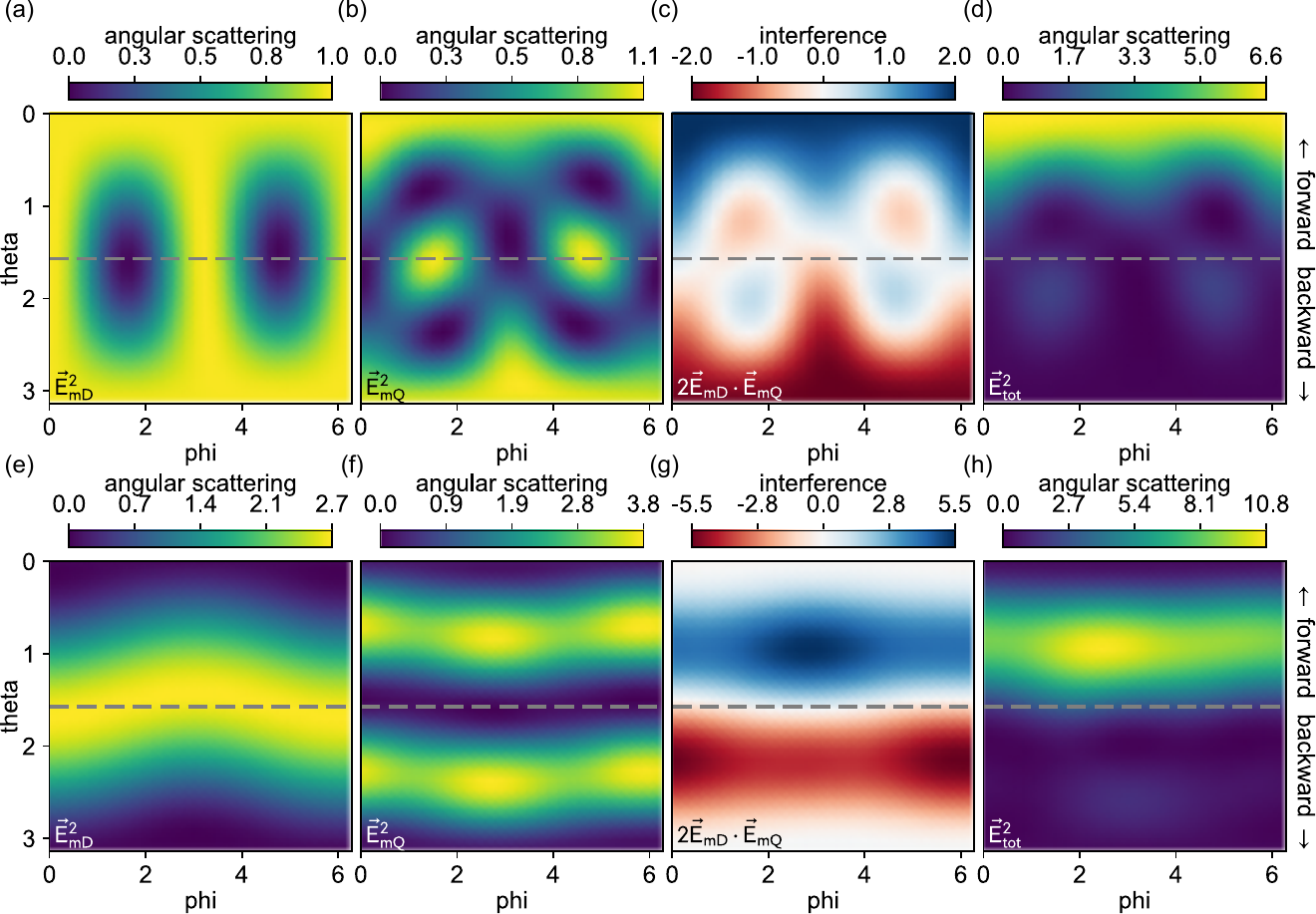}
    \caption{The magnetic dipole and magnetic quadrupole modes spatially interfere affecting the angular scattering distribution.
        (a-d) Angular scattering intensity by linearly polarized beam excitation for a magnetic dipole mode (a), magnetic quadrupole mode (b), magnetic dipole and magnetic quadrupole interference (c), and total scattering (d). 
        (e-h) Corresponding angular scattering intensity by azimuthally polarized beam excitation.
        The dashed line separates the forward and backward scattering regions.
    All images are for a $\SI{1.4}{eV}$ source.}
    \label{fig:interference}
\end{figure}

From Fig. \ref{fig:permutation}, it is apparent that the forward scattering is much stronger than the backward scattering across all permutations for both linearly and azimuthally polarized light.
The dominant scattering modes at low energies are magnetic dipole and magnetic quadrupole modes.
The weak back-scattering can be understood by looking at the interference of these two modes.
Fig. \ref{fig:interference}(a-d) shows the angular distribution of the magnetic dipole, magnetic quadrupole, interference term, and total scattering under linearly polarized beam excitation at \SI{1.4}{eV}.
Each distribution is obtained via the terms in Eq.(\ref{eqn:interference}) and visualized in spherical coordinates.
In Fig. \ref{fig:interference}(c), regions of red correspond to destructive interference while blue regions correspond to constructive interference.
The dashed horizontal line separates the forward scattering ($\theta < \pi/2$) from the backward scattering ($\theta > \pi/2$).

Fig. \ref{fig:interference}(e-h) show the same analysis for azimuthally polarized beam excitation.
Although the dipolar and quadrupolar modes have a different angular distribution than the linearly polarized excitation (see Fig. \ref{fig:setup}(b)), these modes still destructively interfere in the backward direction.
It is important to keep these interference effects in mind when interpreting the results of experiments that measure scattered light in a particular direction.

\begin{figure}[ht!]
    \centering\includegraphics{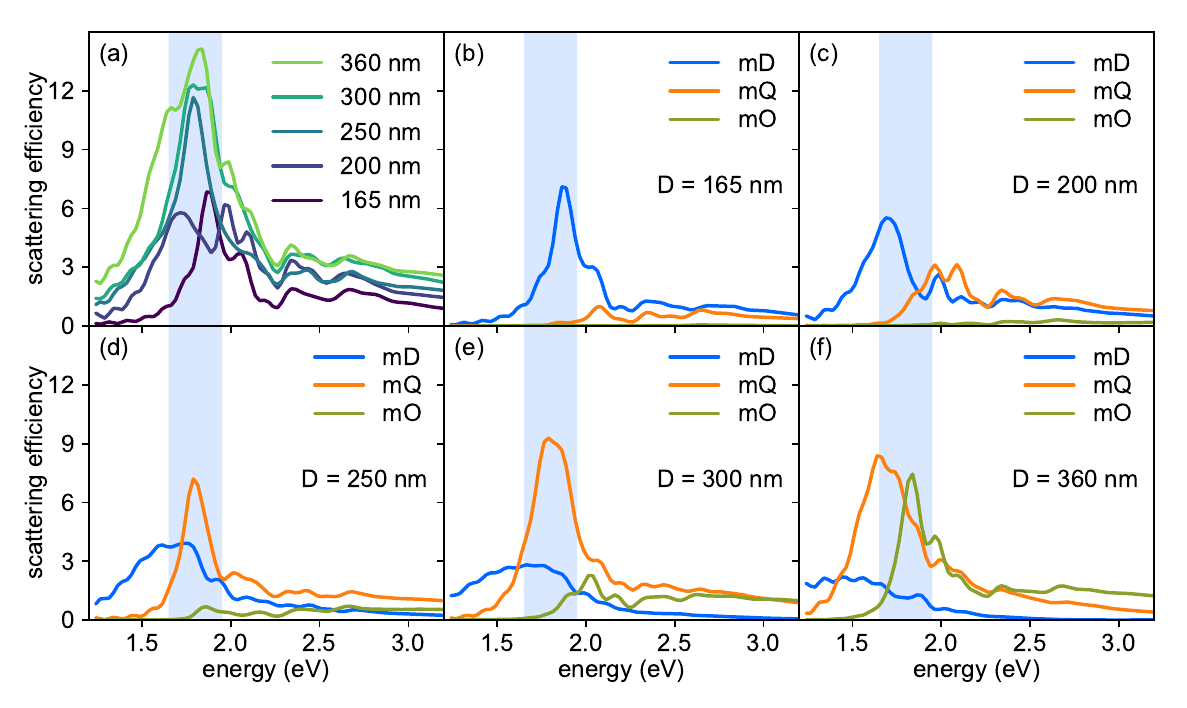}
    \caption{Multipolar analysis with varying core diameter (between $\SI{165}{nm}$ and $\SI{360}{nm}$) for azimuthally polarized beam excitation.
    (a) Total scattering efficiency for each core diameter.
    The magnetic multipolar modes are shifted to lower energies with increasing core diameter, and enhanced in the $\SI{1.7}{eV}$ to $\SI{1.9}{eV}$ blue-shaded band.
    At $D=\SI{165}{nm}$ (b), the magnetic dipole is the main source of scattering, while at $D=\SIrange{200}{300}{nm}$ (c-f) magnetic dipole and magnetic quadrupole modes scatter together.
    The magnetic octupole mode becomes significant at $D=\SI{360}{nm}$ while magnetic dipole mode broadens and continues to red-shift.}
    \label{fig:vary_core}
\end{figure}

The presence of the magnetic quadrupole mode at visible energies is due to the rather large size of the dielectric core.
To better understand these higher order modes, we varied the diameter of the dielectric core from \SIrange{165}{360}{nm}.
The total scattering intensity for each core diameter is shown in Fig. \ref{fig:vary_core}(a) for azimuthally polarized beam excitation.
The collective magnetic spectral feature of the core-satellite structure is present in each structure between \SIrange{1.7}{1.9}{eV} (blue shaded region) with only small spectral shifts.

The multipolar scattering intensities for each of these spectra for azimuthally polarized excitation are is shown in Fig. \ref{fig:vary_core}(b-f).
Only the magnetic modes are depicted owing to the selective behavior of the azimuthally polarized beam (see section \ref{sec:selective}).
For a core diameter of \SI{165}{nm}, only magnetic dipole radiation is present.
We note that for this smaller core size, the interference effects in Fig. \ref{fig:interference} (a core diameter of \SI{330}{nm}) are not present due to a weak magnetic quadrupolar mode.
As the diameter of the core increases from \SIrange{200}{300}{nm}, the magnetic quadrupole mode increases in intensity and overpowers the magnetic dipole.
When $D = \SI{360}{nm}$, magnetic octupole scattering becomes significant.

As expected, these multipolar modes shift to lower energies with increasing core size.
An interesting behavior of these modes is that they spectrally broaden and have weaker (peak) intensity outside of the \SIrange{1.7}{1.9}{eV} energy range, and are enhanced whenever inside this energy range.
Thus, as the magnetic dipole mode red-shifts past \SI{1.7}{eV}, its amplitude weakens, whereas the magnetic quadrupole red-shifts into the \SIrange{1.7}{1.9}{eV} region and its amplitude is enhanced.
This explains the apparent lack of a spectral shift in the total scattering spectra of Fig. \ref{fig:vary_core}(a) as the core size increases.

Size variation of the core diameter was also done for linearly polarized beam excitation.
In this case, the magnetic modes of the system behave similarly to those shown in Fig. \ref{fig:vary_core} with the addition of electric modes at higher energies (data not shown).

\subsection{Selective excitations by vector beams} \label{sec:selective}

\begin{figure}[ht!]
    \centering\includegraphics{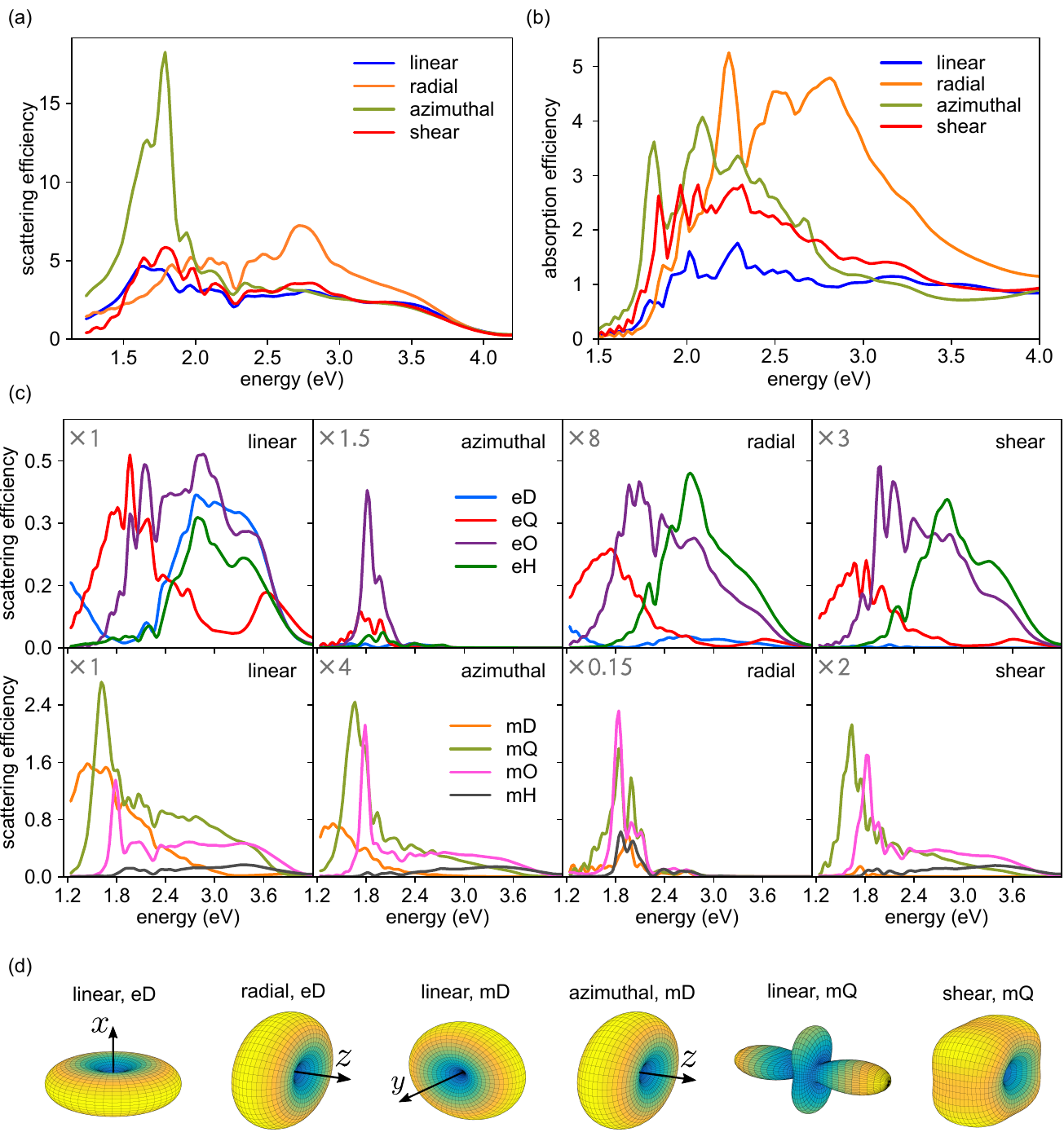}
    \caption{Scalar and vector beams have different selectivity in exciting, enhancing, and orienting the optical modes of the randomly arranged \SI{165}{nm} radius core-satellite structure.
            (a) Total scattering efficiency for each type of beam.
            (b) Absorption efficiency for each type of beam.
    (c) Multipole scattering for each vector beam compared to scattering from a linearly polarized beam.
        The first row shows the electric modes and the second row shows the magnetic modes.
        Each axis has a multiplier to indicate its enhancement relative to the linearly polarized beam excitation, i.e. the azimuthally polarized beam enhances the magnetic dipole by a factor of 2 and the magnetic quadrupole by a factor of 4 over the corresponding linearly polarized beam modes.
        The labels indicate the multipolar mode: e = electric, m = magnetic, D = dipole, Q = quadrupole, O = octupole, H = hexadecapole.
    (d) Angular distribution of $|\boldsymbol{E}(\theta, \phi)|^2$ for different modes and beam excitations.}
    \label{fig:selective}
\end{figure}

Fig. \ref{fig:selective}(a) shows the total (\SI{4\pi}{sr}) scattering spectra of a SiO$_2$-Ag core-satellite structure resulting from excitation by a linearly polarized beam and each of the three vector beams.
Compared to the linearly polarized beam scattering, the azimuthally polarized beam exhibits enhanced scattering at \SI{1.8}{eV} and the radially polarized beam has enhanced scattering at \SI{2.8}{eV}.
The ``shear'' beam has total scattering similar to that of the linearly polarized beam.
These results are found by using a standard box monitor surrounding the core-satellite structure to obtain the total scattered flux (see Eq.(\ref{eqn:box_flux})).

A corresponding absorption spectrum for linearly polarized light and each vector beam is shown in Fig. \ref{fig:selective}(b).
The absorption under vector beam illumination is found to also be enhanced relative to the absorption under linearly polarized beam excitation.

Fig. \ref{fig:selective}(c) shows the corresponding multipolar scattering obtained via the spherical monitor for each of the four beams calculated using Eq.(\ref{eqn:multipole_flux}).
The multipolar scattering of each vector beam excitation from the core-satellite structure is compared to the corresponding modes excited by the linearly polarized beam on the same $y$-scale.
For linearly polarized beam excitation, the magnetic modes (bottom row) are stronger than the electric modes (top row) and occur at lower energies.

The magnetic modes of the radially polarized beam are weaker than the corresponding magnetic modes excited by the linearly polarized beam by a factor of 6.
However, the electric modes of the radially polarized beam are enhanced by up to a factor of 8 relative to the electric modes from linearly polarized beam excitation.
This multipolar analysis leads to the following `selection rule': \emph{the radially polarized beam selectively excites and enhances the electric modes of the system.}

By contrast, \emph{the azimuthally polarized beam selectively excites magnetic modes} and enhances them up to a factor of 7 relative to the magnetic modes vs. linearly polarized beam excitation.
The electric modes are weak under azimuthally polarized light relative to the electric modes excited by linearly polarized light, except in the energy range \SI{1.6}{eV} to \SI{2.2}{eV}, where the electric octupole mode is slightly enhanced.
These enhancement factors for radially and azimuthally polarized beam excitations explain the corresponding enhancements in the total scattering in Fig. \ref{fig:selective}(a) at \SI{1.8}{eV} and \SI{2.8}{eV}. 

For excitation with a shear polarized beam, both electric and magnetic modes are found to be enhanced vs. the modes from linearly polarized beam excitation, except the modes of dipolar order.
Thus, \emph{the shear polarized beam selectively excites and enhances modes of quadrupole order or higher, both electric and magnetic modes}.
The enhancement factor of these modes is around 2 to 3, somewhat smaller than the enhancement factors from azimuthally and radially polarized beams.

Fig. \ref{fig:selective}(d) shows the angular distribution of the scattering intensity for different beams and multipolar modes.
The linearly $x$-polarized beam excites an $x$-oriented electric dipole and $y$-oriented magnetic dipole.
The electric dipole selectively excited by the radially polarized beam is oriented along the $z$-axis (the beam axis).
Similarly, the magnetic dipole selectively excited by the azimuthally polarized beam is oriented along the beam axis.
The associated distributions of scattered radiation can have important effects in experiments that only measure scattering intensity over a finite collection angle in a particular direction (e.g. forward, backward, perpendicular to the beam axis, etc.).

The magnetic quadrupole mode is also shown to be of a different nature in the case of shear polarized beam excitation.
Any quadrupole mode can be expressed as a sum of 5 distinct quadrupole modes, denoted by $m=-2,-1,0,1,2$ (see Eq.(\ref{eqn:vcs_expansion})).
Different values of $m$ correspond to different angular distributions of the quadrupole mode.
The quadrupole mode excited by the linearly polarized beam contains equal parts of $m=-1$ and $m=1$, while the mode excited by the shear polarized beam contains equal parts of $m=-2$ and $m=2$.
The magnetic quadrupole mode excited by the azimuthally polarized beam (not shown) only contains the mode with $m=0$.
Thus, each of the beams selectively excites a different kind of magnetic quadrupole.
Similar results are obtained for the electric quadrupole mode excited by the linear, radial, and shear polarized beams.

\section{Concluding remarks}

We have introduced a method of multipolar analysis in FDTD simulations and applied it to understand the interaction of vector beams with SiO$_2$-Ag core-satellite structures at a new level of detail.
This method can be used to identify the multipolar nature of scattering peaks and provide angular scattering intensity for better comparison to experimental results.
In addition, nanoparticle placement and arrangement is shown to be an important aspect of scattering by these systems, which must be considered when investigating metamaterials and metafluids composed of random ensembles.

FDTD simulations revealed the nature of the magnetic dipole mode to be a circulation of displacement currents around the perimeter of the core-satellite structure.
For linearly polarized light, the displacement currents circulate in the $xz$ plane while for azimuthally polarized light, they circulate in the $xy$ plane.
These two magnetic modes are excited by different mechanisms: in the case of linearly polarized light, retardation leads to a circulation of the fields while for azimuthally polarized light, the unique polarization state of the electric field drives the circulation.
Exploration of the consequences of these resonances on the effective electric permittivity and magnetic permeability in both individual and ensembles of core-shell systems may prove to be very interesting from the point of view of developing novel metameterials.


One future direction of study is to better understand and visualize the quadrupolar and higher order multipolar modes excited by the vector beams as has been done here for the magnetic dipolar mode (see Figs. \ref{fig:magnetic_mode_2}, \ref{fig:field_enhancements}).
We determined that the quadrupolar mode plays an important role in the angular scattering intensity of larger core-satellite nanostructures through interference with the dipolar modes resulting in destructive interference in the backward scattering direction and constructive interference in the forward scattering direction.
These multipolar interference patterns are important when considering experiments that collect over a finite collection angle, and also have applications in altering and guiding the scattered light field.

The multipolar analysis we introduced combined with vector beam excitation of core-satellite structures reveals a class of selection rules:
(1) azimuthally polarized light is shown to selectively excite the magnetic modes of the core-satellite structure, in addition to enhancing their scattering intensity and spatially rotating the mode vs. the magnetic mode excited by a linearly polarized beam,
(2) radially polarized light selectively excites and enhances the electric modes vs. linearly polarized excitation,
(3) shear polarized light selectively excites and enhances electric and magnetic quadrupolar and higher order modes.
These selection rules help identify and quantify the scattering properties of individual nanostructures, and will be helpful in the design of metamaterials and metafluids.
A generalization to higher order vector beams may also lead to a broader set of selection rules, allowing fine-tuned control over the scattering properties of individual nanostructures.

\section*{Funding}


Vannevar Bush Faculty Fellowship of the Office of Naval Research (N00014-16-1-2502);
Center for Nanoscale Materials of Argonne National Laboratory (DE-AC02-06CH11357).
\section*{Acknowledgments}

The authors acknowledge support from the Vannevar Bush Faculty Fellowship program sponsored by the Basic Research Office of the Assistant Secretary of Defense for Research and Engineering and funded by the Office of Naval Research.
This work was performed, in part, at the Center for Nanoscale Materials, a U.S. Department of Energy Office of Science User Facility.
We also acknowledge the University of Chicago Research Computing Center for providing the computational resources needed for this work. \newline

\bibliographystyle{unsrt}
\bibliography{manuscript.bib}{}

\begin{thebibliography}{10}

\bibitem{veselago1968electrodynamics}
V.~G. Veselago.
\newblock The electrodynamics of substances with simultaneously negative values
  of $\epsilon$ and $\mu$.
\newblock {\em Soviet Physics Uspekhi}, 10(4):509, 1968.

\bibitem{smith2004metamaterials}
D.~R. Smith, J.~B. Pendry, and MCK Wiltshire.
\newblock Metamaterials and negative refractive index.
\newblock {\em Science}, 305(5685):788--792, 2004.

\bibitem{bourgeois2017self}
M.~R. Bourgeois, A.~T. Liu, M.~B. Ross, J.~M. Berlin, and G.~C. Schatz.
\newblock Self-assembled plasmonic metamolecules exhibiting tunable magnetic
  response at optical frequencies.
\newblock {\em The Journal of Physical Chemistry C}, 121(29):15915--15921,
  2017.

\bibitem{fan2010self}
J.~A. Fan, C.~Wu, K.~Bao, J.~Bao, R.~Bardhan, N.~J. Halas, V.~N. Manoharan,
  P.~Nordlander, G.~Shvets, and F.~Capasso.
\newblock Self-assembled plasmonic nanoparticle clusters.
\newblock {\em Science}, 328(5982):1135--1138, 2010.

\bibitem{alu2006negative}
A.~Al{\`u}, A.~Salandrino, and N.~Engheta.
\newblock Negative effective permeability and left-handed materials at optical
  frequencies.
\newblock {\em Optics Express}, 14(4):1557--1567, 2006.

\bibitem{sheikholeslami2013metafluid}
S.~N. Sheikholeslami, H.~Alaeian, A.~L. Koh, and J.~A. Dionne.
\newblock A metafluid exhibiting strong optical magnetism.
\newblock {\em Nano Letters}, 13(9):4137--4141, 2013.

\bibitem{fruhnert2014towards}
M.~Fruhnert, S.~M{\"u}hlig, F.~Lederer, and C.~Rockstuhl.
\newblock Towards negative index self-assembled metamaterials.
\newblock {\em Physical Review B}.

\bibitem{qian2015raspberry}
Z.~Qian, S.~P. Hastings, C.~Li, B.~Edward, C.~K. McGinn, N.~Engheta,
  Z.~Fakhraai, and S.~Park.
\newblock Raspberry-like metamolecules exhibiting strong magnetic resonances.
\newblock {\em ACS Nano}, 9(2):1263--1270, 2015.

\bibitem{vallecchi2011collective}
A.~Vallecchi, M.~Albani, and F.~Capolino.
\newblock Collective electric and magnetic plasmonic resonances in spherical
  nanoclusters.
\newblock {\em Optics Express}, 19(3):2754--2772, 2011.

\bibitem{muhlig2011self}
S.~M{\"u}hlig, A.~Cunningham, S.~Scheeler, C.~Pacholski, T.~B{\"u}rgi,
  C.~Rockstuhl, and F.~Lederer.
\newblock Self-assembled plasmonic core--shell clusters with an isotropic
  magnetic dipole response in the visible range.
\newblock {\em ACS Nano}, 5(8):6586--6592, 2011.

\bibitem{ponsinet2015resonant}
V.~Ponsinet, P.~Barois, S.~M. Gali, P.~Richetti, J.~Salmon, A.~Vallecchi,
  M.~Albani, A.~Le~Beulze, S.~Gomez-Grana, E.~Duguet, S.~Mornet, and
  M.~Treguer-Delapierre.
\newblock Resonant isotropic optical magnetism of plasmonic nanoclusters in
  visible light.
\newblock {\em Physical Review B}, 92(22):220414, 2015.

\bibitem{gandra2012plasmonic}
N.~Gandra, A.~Abbas, L.~Tian, and S.~Singamaneni.
\newblock Plasmonic planet--satellite analogues: hierarchical self-assembly of
  gold nanostructures.
\newblock {\em Nano Letters}, 12(5):2645--2651, 2012.

\bibitem{uttam}
U.~Manna, J.~Lee, T.~Deng, J.~Parker, N.~Shepherd, Y.~Weizmann, and N.~F.
  Scherer.
\newblock Selective induction of optical magnetism.
\newblock {\em Nano Letters}, 2017.

\bibitem{das2015beam}
T.~Das, P.~P. Iyer, R.~A. DeCrescent, and J.~A. Schuller.
\newblock Beam engineering for selective and enhanced coupling to multipolar
  resonances.
\newblock {\em Physical Review B}, 92(24):241110, 2015.

\bibitem{wozniak2015selective}
P.~Wo{\'z}niak, P.~Banzer, and G.~Leuchs.
\newblock Selective switching of individual multipole resonances in single
  dielectric nanoparticles.
\newblock {\em Laser \& Photonics Reviews}, 9(2):231--240, 2015.

\bibitem{hancu2013multipolar}
I.~M. Hancu, A.~G. Curto, M.~Castro-L{\'o}pez, M.~Kuttge, and N.~F. van Hulst.
\newblock Multipolar interference for directed light emission.
\newblock {\em Nano Letters}, 14(1):166--171, 2013.

\bibitem{kujala2007multipole}
S.~Kujala, B.~K. Canfield, M.~Kauranen, Y.~Svirko, and J.~Turunen.
\newblock Multipole interference in the second-harmonic optical radiation from
  gold nanoparticles.
\newblock {\em Physical Review Letters}, 98(16):167403, 2007.

\bibitem{oldenburg1999light}
S.~J. Oldenburg, G.~D. Hale, J.~B. Jackson, and N.~J. Halas.
\newblock Light scattering from dipole and quadrupole nanoshell antennas.
\newblock {\em Applied Physics Letters}, 75(8):1063--1065, 1999.

\bibitem{taflove1995computational}
A.~Taflove and S.~C. Hagness.
\newblock Computational electrodynamics: the finite-difference time-domain
  method.
\newblock {\em Norwood, 2nd Edition, MA: Artech House}, 1995.

\bibitem{OskooiRo10}
F.~O. Ardavan, R.~David, I.~Mihai, B.~Peter, J.~D. Joannopoulos, and G.~J.
  Steven.
\newblock {MEEP}: A flexible free-software package for electromagnetic
  simulations by the {FDTD} method.
\newblock {\em Computer Physics Communications}, 181:687--702, January 2010.

\bibitem{FarjadpourRo06}
A.~Farjadpour, D.~Roundy, A.~Rodriguez, M.~Ibanescu, P.~Bermel, J.~D.
  Joannopoulos, S.~G. Johnson, and G.~Burr.
\newblock Improving accuracy by subpixel smoothing in {FDTD}.
\newblock {\em Optics Letters}, 31:2972--2974, October 2006.

\bibitem{OskooiKo09}
A.~F. Oskooi, C.~Kottke, and S.~G. Johnson.
\newblock Accurate finite-difference time-domain simulation of anisotropic
  media by subpixel smoothing.
\newblock {\em Optics Letters}, 34:2778--2780, September 2009.

\bibitem{johnson1972optical}
P.~B. Johnson and R.~W. Christy.
\newblock Optical constants of the noble metals.
\newblock {\em Physical Review B}, 6(12):4370, 1972.

\bibitem{rakic1998optical}
A.~D. Raki{\'c}, A.~B. Djuri{\v{s}}i{\'c}, J.~M. Elazar, and M.~L. Majewski.
\newblock Optical properties of metallic films for vertical-cavity
  optoelectronic devices.
\newblock {\em Applied Optics}, 37(22):5271--5283, 1998.

\bibitem{zhan2009cylindrical}
Q.~Zhan.
\newblock Cylindrical vector beams: from mathematical concepts to applications.
\newblock {\em Advances in Optics and Photonics}, 1(1):1--57, 2009.

\bibitem{novotny2012principles}
L.~Novotny and B.~Hecht.
\newblock {\em Principles of Nano-Optics}.
\newblock Cambridge University Press, 2012.

\bibitem{rengarajan2000field}
S.~R. Rengarajan and Y.~Rahmat-Samii.
\newblock The field equivalence principle: Illustration of the establishment of
  the non-intuitive null fields.
\newblock {\em IEEE Antennas and Propagation Magazine}, 42(4):122--128, 2000.

\bibitem{muhlig2011multipole}
S.~M{\"u}hlig, C.~Menzel, C.~Rockstuhl, and F.~Lederer.
\newblock Multipole analysis of meta-atoms.
\newblock {\em Metamaterials}, 5(2):64--73, 2011.

\bibitem{grahn2012electromagnetic}
P.~Grahn, A.~Shevchenko, and M.~Kaivola.
\newblock Electromagnetic multipole theory for optical nanomaterials.
\newblock {\em New Journal of Physics}, 14(9):093033, 2012.

\end{thebibliography}

\end{document}